# MUON COLLIDER INTERACTION REGION DESIGN*

Y.I. Alexahin#, E. Gianfelice-Wendt, V. V. Kashikhin, N.V. Mokhov, A.V. Zlobin, FNAL, Batavia IL, 60510 USA

V.Y. Alexahin, JINR, Dubna, 141980 Russia.

## Abstract

Design of a muon collider interaction region (IR) presents a number of challenges arising from low $\beta^*$ < 1 cm, correspondingly large beta-function values and beam sizes at IR magnets, as well as the necessity to protect superconducting magnets and collider detectors from muon decay products. As a consequence, the designs of the IR optics, magnets and machine-detector interface are strongly interlaced and iterative. A consistent solution for the 1.5 TeV c.o.m. muon collider IR is presented. It can provide an average luminosity of $10^{34}$ cm$^{-2}$s$^{-1}$ with an adequate protection of magnet and detector components.

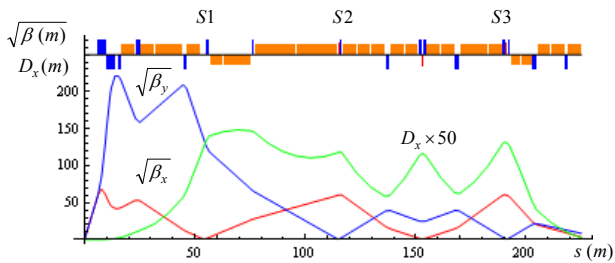

Figure 1 (color): IR layout and optics functions.

# INTRODUCTION

To satisfy requirements on a high-luminosity muon collider lattice, a new approach to the IR chromaticity correction was developed [1]. The IR layout and optics functions for $\beta^*$ = 1 cm are shown in Fig. 1. Dipoles (shown at the top as orange rectangles) are placed immediately after the Final Focus (FF) doublet and generate a sufficiently large dispersion function at the location of the nearest to the interaction point (IP) sextupole S1 which compensates vertical chromaticity.

This design raises a number of questions: large values of vertical $\beta$-function and therefore of the vertical beam-size in the IR quads and dipoles make it necessary to reconsider earlier magnet designs, closeness of the dipoles to IP may complicate the detector protection from $\gamma$-radiation emitted by decay electrons and positrons and from these electrons and positrons themselves.

These issues as well as problems with heat deposition in the magnet coils are considered in the present report.

# IR MAGNET DESIGN

Fig. 2 shows beam sizes corresponding to the muon beam parameters: $E$ = 0.75 TeV, $\varepsilon_{LN}$ = 25 $\pi$·mm·mrad,

* Work supported by Fermi Research Alliance, LLC under Contract DE-AC02-07CH11359 with the U.S. DOE.

# alexahin@fnal.gov

$\sigma_p/p$ = 0.1% and the inner radii of closest to IP magnets determined by the requirement $a > 5\sigma_{max}$+1 cm. In the actual magnet design, the bore radius was increased by additional 5 mm to provide an adequate space for the beam pipe, annular helium channel and possible inner absorber (liner).

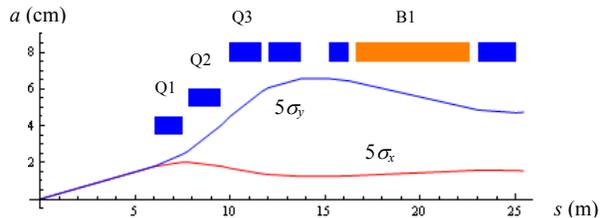

Figure 2: Beam sizes and aperture of the FF magnets.

## FF Quadrupoles

The IR quadrupoles are split in pieces of no more than 2 m long to allow for placement of protecting tungsten masks between them. The first two quadrupoles in Fig. 2 are focusing ones and the next three are defocusing ones. The space between the 4th and 5th quadrupoles is reserved for beam diagnostics and correctors.

The requirements to the FF quadrupoles are close to those being considered for the LHC luminosity upgrade [2], allowing one to use their design concepts for the MC IR. The MC IR quadrupoles based on Nb$_3$Sn superconductor are described in [3] and their parameters summarized in Table 1. As can be seen, all the magnets have ~12% margin at 4.5 K, which is sufficient for the stable operation with the average heat deposition in magnet mid-planes up to 1.7 mW/g. Operation at 1.9 K would increase the magnet margin to ~22% and their quench limit by a factor of 4.

Table 1: IR quadrupole parameters.

| Parameter | Unit | Q1 | Q2 | Q3 |
|---|---|---|---|---|
| Coil aperture | mm | 80 | 110 | 160 |
| Nominal gradient | T/m | 250 | 187 | -130 |
| Nominal current | kA | 16.61 | 15.3 | 14.2 |
| Quench gradient @ 4.5 K | T/m | 281.5 | 209.0 | 146.0 |
| Quench gradient @ 1.9 K | T/m | 307.6 | 228.4 | 159.5 |

## Open Midplane Dipole

The vertical elongation of the beam makes requirements to the IR dipoles quite different from those to the arc dipoles where the horizontal aperture must be larger due to the orbit sagitta and large dispersion

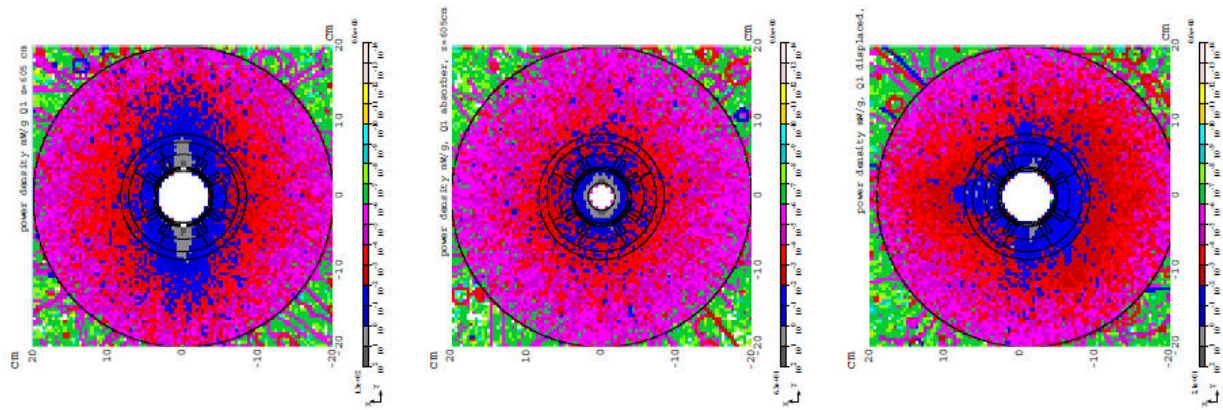

Figure 3 (color): Deposited power density in Q1 (mW/g) for three cases: "standard" (left), with absorbers inside (center) and with horizontal displacement (right). Larger radii are on the left of the plots.

contribution to the beam size. Also, it is even more important for the IR dipoles to have an open mid-plane to avoid showering of energetic decay electrons in a vicinity of the superconducting coils and absorb them in high-Z rods cooled at liquid nitrogen temperatures and placed far from the coils. This has also a potential of reducing background fluxes in a central tracker. To remove 95% of radiation the full gap between the poles should be at least $5\sigma_y$ or 6 cm. This large gap limits the bending field which can be achieved with $Nb_3Sn$ coils and make it more difficult to achieve an acceptable field quality in the required aperture.

Several options were considered for such a dipole based on $Nb_3Sn$ superconductor with the required bending field of 8 T, good field quality in the aperture with 100 mm in vertical direction and 50 mm in horizontal direction, and large margin at 4.5 K [3]. Parameters of the open-midplane IR dipole version used in this analysis are reported in Table 2. More detail is given in [3].

Table 2: IR dipole parameters.

| Coil aperture | mm | 160 |
|---|---|---|
| Gap | mm | 55 |
| Nominal field | T | 8 |
| Nominal current | kA | 17.85 |
| Quench field @ 4.5 K | T | 9.82 |

## ENERGY DEPOSITION IN IR MAGNETS

Energy deposition and detector backgrounds are simulated with the MARS15 code [4]. All the related details of geometry, materials distributions and magnetic fields are implemented into the model for lattice elements and tunnel in the ±200-m from IP region, 4th concept detector components [5], experimental hall and machine-detector interface. To protect SC magnets and detector, tungsten masks in the interconnect regions, liners in magnet apertures (wherever needed), and a sophisticated tungsten cone inside the detector were implemented into the model and carefully optimized. The muon beam energy assumed in this study is 750 GeV, with 2.e12 muons per bunch and 15 Hz repetition rate. The muon beam is aborted after 1000 turns when the luminosity is reduced by a factor of 3.

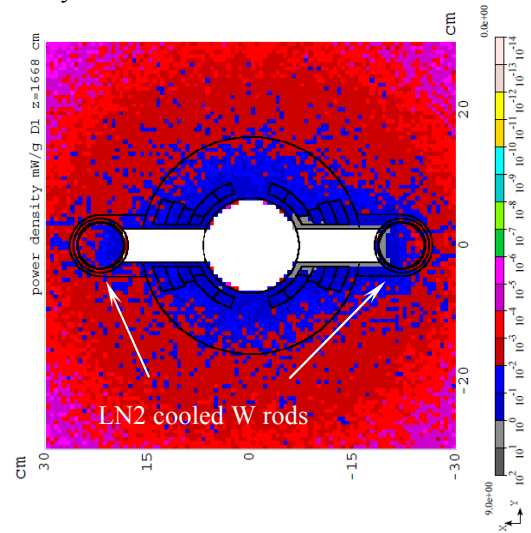

Figure 4 (color): Power density (mW/g) in B1 dipole for case (iii).

Three cases were considered: (i) "standard" when 10-cm thick tungsten masks with 5 $\sigma_{x,y}$ elliptic openings are put in the IR magnet interconnect regions; (ii) with additional tungsten liners inside the quadrupoles leaving a 5 $\sigma_{x,y}$ elliptic aperture for the beam; (iii) as first case, but with the IR quadrupoles displaced horizontally by 0.1 of their apertures, so as to provide ~2 T bending field. This additional field helps also facilitate chromaticity correction by increasing dispersion at the sextupoles, and deflect low-energy charged particles from the detector.

Power density isocontours at shower maximum in the first quadrupole are shown in Fig. 3, while Fig. 4 displays such profiles in the IR dipole B1. Maximum values of power density in the most vulnerable magnets are presented in Table 3. One can see that quadrupole displacement reduces power density but not enough to avoid using liners inside quadrupoles. Combining all the three cases has a potential of keeping peak power density in the IR magnets below the quench limits.

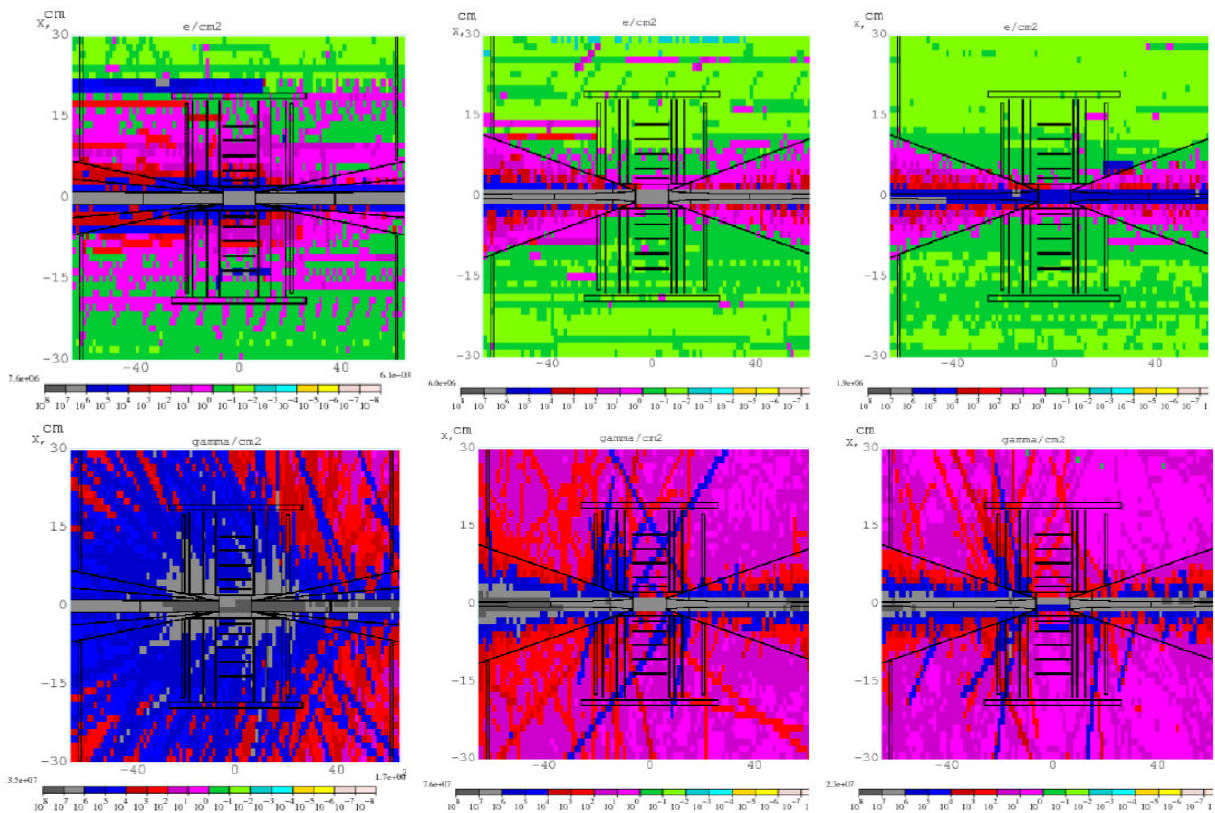

Figure 5 (color): Electron (top) and gamma (bottom) fluxes in the detector in three cases described in the text.

Table 3: Peak power density (mW/g) in considered cases.

| Magnet | (i) | (ii) | (iii) |
|--------|-----|------|-------|
| Q1 | 5.0 | 1.0 | 3.0 |
| Q2 | 10. | 1.0 | 10. |
| Q5 | 3.7 | 2.0 | 3.7 |
| B1 | 3.0 | 2.6 | 1.9 |
| Q6 | 3.6 | 2.6 | 2.0 |

## DETECTOR BACKGROUNDS

Figure 5 compares calculated electron and gamma fluxes for the following cases: left – no masks between magnets, 6° cone with a 5σ radius liner up to 2 m from IP; center - 5σ masks inserted between FF quads, cone angle increased to 10°, 5σ liner up to 1 m from IP; right – same as above plus FF quad displacement.

The masks and increased cone angle reduce the electron and gamma fluxes by factors 300 and 20 respectively. Displacing the FF quads slightly increases the electron flux (by ~50%) but decreases the gamma flux by another factor of 15, so the overall effect of quad displacement may be considered as positive.

Results of further optimization of the cone nose geometry are presented in Fig. 6. It shows gamma flux as a function of the angle of inner cone opening towards IP at the outer cone angle of 10°.

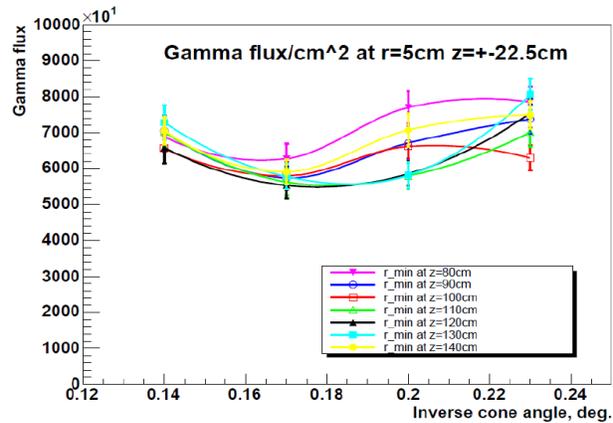

Figure 6 (color): Gamma flux vs. inner cone angle at different positions of minimal aperture from IP